\documentclass[aps,showpacs,twocolumn]{revtex4}

\usepackage{graphicx}
\usepackage{epstopdf}
\usepackage{dcolumn}
\usepackage[ulem=normalem]{changes}

\begin{document}
\title{Geometry of the shears mechanism in nuclei}
\author{P.~Van~Isacker$^1$}
\author{A.O.~Macchiavelli$^2$}
\affiliation{
$^1$Grand Acc\'el\'erateur National d'Ions Lourds,
CEA/DSM--CNRS/IN2P3, B.P.~55027, F-14076 Caen Cedex 5, France}
\affiliation{
$^2$Nuclear Science Division, Lawrence Berkeley National Laboratory,
Berkeley, California 94720}
\date{\today}

\begin{abstract}
The geometry of the shears mechanism in nuclei
is derived from the nuclear shell model.
This is achieved by taking the limit of large angular momenta
(classical limit) of shell-model matrix elements.
\end{abstract}
\maketitle

A central theme in the study of quantum many-body systems
is the understanding of its elementary modes of excitation.  
Of particular interest is the competition
between single-particle and collective degrees of freedom.
In general, single-particle motion gives rise to irregular level sequences,
associated with the unique properties of the individual particles.
Collective motion generates more regular structures,
a typical example being that of rotational bands
which follow closely a $J(J+1)$ sequence,
characteristic of a quantum rotor.

In atomic nuclei, the energy scales of these modes are somewhat comparable,
leading to a rich and complex structure~\cite{BM75}.
$^{208}$Pb is considered a paradigm of a doubly-closed-shell nucleus
with collective states based on an octupole phonon.
Neighboring nuclei provided a wealth of important information for the shell model,
on single-particle levels and residual interactions at the $Z=82$, $N=126$ shell closure,
as well as on the role of particle-vibration coupling~\cite{Hamamoto74}.

The observation in $^{199}$Pb of a regular sequence of $\gamma$ rays
resembling, at first look, a rotational band
came as a surprise~\cite{Baldsiefen92,Clark92,Clark00,Hubel05}. 
A closer inspection of the data established that these transitions were of magnetic character,
in contrast to the quadrupole nature in the well-known nuclear rotors \cite{BM75}.
The answer to this puzzle was provided by Frauendorf~\cite{Frauendorf93},
who based on the Titled Axis Cranking Model proposed the so-called ``shears mechanism".
He showed that for a weakly deformed system,
there exist low-energy configurations in which neutrons and protons
combine into stretched structures (``blades")
and the angular momentum is generated by the re-coupling of these blades,
resembling the closing of a pair of shears. 
As discussed in Ref.~\cite{Frauendorf01},
the breaking of the rotational symmetry in this case originates in an anisotropic
distribution of nucleonic current loops (rather than electric charge),
and thus it is usually referred as magnetic rotation.
Further experiments not only confirmed this picture in the lead region,
but also established the mechanism in other regions
of the Segr\'e chart,
near doubly-magic closures. 

While an interpretation in terms of the shears mechanism
provides an appealing, intuitive picture of  these nuclear states,
the question remains whether
this geometric character is borne out by microscopic calculations.
In numerical shell-model calculations
Frauendorf {\it et al.}~\cite{Frauendorf96} showed
that the shears picture is valid in the lead isotopes.
A direct derivation of the geometry of the shears mechanism from the shell model
has, however, to our knowledge never been given.
Providing such a derivation
is the purpose of the present Letter.

Let us assume that two nucleons of one type (say neutrons)
occupy particle-like orbits $j_{1\nu}$ and $j_{2\nu}$,
while two nucleons of the other type (protons)
occupy hole-like orbits $j_{1\pi}^{-1}$ and $j_{1\pi}^{-1}$~\cite{note1}.
The neutrons and protons
are coupled to angular momenta $J_\nu$ and $J_\pi$, respectively,
which are close to stretched, $J_\rho\approx j_{1\rho}+j_{2\rho}$ ($\rho=\nu,\pi$).
The two-particle (2p) and two-hole (2h) states
are represented as
$|N\rangle=|j_{1\nu}j_{2\nu};J_\nu\rangle$ and 
$|P^{-1}\rangle=|j_{1\pi}^{-1}j_{2\pi}^{-1};J_\pi\rangle$,
and we assume that the states are Pauli allowed,
{\it i.e.}, that $J_\rho$ is even if $j_{1\rho}=j_{2\rho}$.
A shears band consists of the states $|NP^{-1};J\rangle$
where $J$ results from the coupling of $J_\nu$ and $J_\pi$.

We ask the following questions:
How do the energies of the members of this band evolve 
as a function of $J$,
and how does this evolution depends
on the angular momenta of the single-particle orbits
and the angular momenta of the blades?
We answer these questions by adopting a shell-model hamiltonian
of the generic form
$H=H_\nu+H_\pi+V_{\nu\pi}$
and computing $\langle NP^{-1};J|H|NP^{-1};J\rangle$,
which will be referred to
as the (2p-2h) shears matrix element.

The shell-model hamiltonian is specified by
the single-particle energies,
the single-hole energies,
the neutron-neutron and proton-proton (two-body) interaction matrix elements,
and the neutron-proton (np) interaction matrix elements
$V_{j_\nu j_\pi}^R\equiv\langle j_\nu j_\pi;R|V_{\nu\pi}|j_\nu j_\pi;R\rangle$.
One finds that the energy contribution of $H_\nu$ and $H_\pi$
is constant for all members of the shears band
and that any $J$ dependence originates from the np interaction $V_{\nu\pi}$.
A multipole expansion of the latter interaction
leads to the following expression for the shears matrix element:
\begin{eqnarray}
\lefteqn{
\langle NP^{-1};J|V_{\nu\pi}|NP^{-1};J\rangle}
\nonumber\\&=&
-{\cal P}\hat J_\nu\hat J_\pi
\sum_R\hat R\,V_{j_{1\nu}j_{1\pi}}^R
\left\{\begin{array}{cccccccc}
\!\!j_{1\nu}\!\!\!\!&&\!\!J_\nu\!\!&&\!\!J_\pi\!\!&&\!\!j_{1\pi}\!\!\!\!&\\
&\!\!j_{2\nu}\!\!\!\!&&\!\!J\!\!&&\!\!j_{2\pi}\!\!\!\!&&\!\!R\!\!\\
\!\!j_{1\nu}\!\!\!\!&&\!\!J_\nu\!\!&&\!\!J_\pi\!\!&&\!\!j_{1\pi}\!\!\!\!&
\end{array}\right\},
\label{e_result1}
\end{eqnarray}
where $\hat x\equiv 2x+1$
and ${\cal P}\equiv{\cal P}_\nu{\cal P}_\pi$
with ${\cal P}_\rho$ an operator
defined from ${\cal P}_\rho f(j_{1\rho},j_{2\rho})=f(j_{1\rho},j_{2\rho})+f(j_{2\rho},j_{1\rho})$
for any function $f$.
The object in curly brackets is a 12$j$ symbol of the {\em first} kind,
a quantity which is scalar under rotations and depends on twelve angular momenta.
The result~(\ref{e_result1}) is obtained
by expressing the shears matrix element
as a sum over four 6$j$ symbols,
which can be related to a 12$j$ symbol,
see Eq.~(19.1) of Ref.~\cite{Yutsis62}.

It is now a simple matter
to introduce in the sum~(\ref{e_result1})
values for the np interaction matrix elements
and to derive the $J$ dependence of the shears matrix element.
For any reasonable nuclear interaction
and for $J_\rho\approx j_{1\rho}+j_{2\rho}$,
it is found that the sum~(\ref{e_result1})
has an approximate parabolic behavior
around its minimum around $J^2\approx J_\nu^2+J_\pi^2$.

To understand better this numerical finding,
the geometry of the expression~(\ref{e_result1}) can be studied
by taking the limit of large angular momenta in the recoupling coefficients.
Such limits are known 
since the seminal study of Wigner
on the classical limit of $3j$ and $6j$ symbols
(see chapter~27 of Ref.~\cite{Wigner59}),
subsequently refined by Ponzano and Regge~\cite{Ponzano68}
whose work was put on a mathematically solid footing
by Schulten and Gordon~\cite{Schulten75}.
The classical limit of a $3j$ symbol
is associated with the area of a (projected) triangle 
while that of a $6j$ symbol
involves the volume of a tetrahedron,
with the lengths of the sides determined by the angular momenta.
Classical limits not only yield approximate expressions
for (re)coupling coefficients
but in addition provide an insight into their geometrical significance.
The study of the classical limit of $3nj$ symbols with $n>2$
is still a topic of active research
with ramifications in fields as diverse as quantum gravity and quantum computing
(see, {\it e.g.} Refs.~\cite{Anderson08,Anderson09,Haggard10}
and references therein),
well beyond the standard applications
in, for example, atomic or nuclear physics.

Since at this moment only partial results are known
for $9j$ (let alone $12j$) symbols,
a classical limit of the expression~(\ref{e_result1})
for arbitrary interactions is difficult to obtain.
We consider instead the surface delta interaction (SDI)~\cite{Brussaard77},
$V^{\rm SDI}(i,j)=-4\pi a'_T\delta(\bar r_i-\bar r_j)\delta(r_i-R)$,
which is known to be a reasonable approximation
to the nucleon-nucleon interaction~\cite{note2}.
The np matrix element $V_{j_\nu j_\pi}^R$ of the SDI is~\cite{Brussaard77}
\begin{equation}
-\frac{\hat\jmath_\nu\hat\jmath_\pi}{2}
\left[a_{01}\left(\begin{array}{ccc}
j_\nu&j_\pi&R\\{\frac12}&-{\frac12}&0
\end{array}\right)^2+
a_0\left(\begin{array}{ccc}
j_\nu&j_\pi&R\\{\frac12}&{\frac12}&-1
\end{array}\right)^2\right],
\label{e_mesdi}
\end{equation}
with $a_{01}=(a_0+a_1)/2-(-)^{\ell_\nu+\ell_\pi+R}(a_0-a_1)/2$
in terms of the strengths $a_T=a'_TC(R_0)$,
where $C(R_0)$ equals
$R_{n_\nu\ell_\nu}^4(R_0)R_0^2=R_{n_\pi\ell_\pi}^4(R_0)R_0^2$.

When the SDI np matrix elements~(\ref{e_mesdi})
are introduced in the shears matrix element~(\ref{e_result1}),
one encounters sums of the type
\begin{widetext}
\begin{equation}
\sigma_n^{(\lambda)}\equiv
(2J_\nu+1)(2J_\pi+1)
\sum_R(-)^{\lambda R}(2R+1)
\left(\begin{array}{ccc}
j_{1\nu}&j_{1\pi}&R\\{\frac12}&n-{\frac12}&-n
\end{array}\right)^2
\left\{\begin{array}{cccccccc}
\!\!j_{1\nu}\!\!\!\!&&\!\!J_\nu\!\!&&\!\!J_\pi\!\!&&\!\!j_{1\pi}\!\!\!\!&\\
&\!\!j_{2\nu}\!\!\!\!&&\!\!J\!\!&&\!\!j_{2\pi}\!\!\!\!&&\!\!R\!\!\\
\!\!j_{1\nu}\!\!\!\!&&\!\!J_\nu\!\!&&\!\!J_\pi\!\!&&\!\!j_{1\pi}\!\!\!\!&
\end{array}\right\},
\label{e_sum1}
\end{equation}
for $(\lambda,n)=(0,0)$, (0,1) and (1,0).
These reduce to simple expressions in the classical limit.
For example, for $\lambda=0$,
the sum can be exactly rewritten as
\begin{equation}
\sigma_n^{(0)}=
(2J_\nu+1)(2J_\pi+1)
\sum_{\stackrel{\scriptstyle m_\nu M_\nu}{m_\pi M_\pi}}
\left(\begin{array}{ccc}
j_{1\nu}&j_{2\nu}&J_\nu\\{\frac12}&m_\nu&M_\nu
\end{array}\right)^2
\left(\begin{array}{ccc}
j_{1\pi}&j_{2\pi}&J_\pi\\-n+{\frac12}&m_\pi&M_\pi
\end{array}\right)^2
\left(\begin{array}{ccc}
J_\nu&J_\pi&J\\M_\nu&M_\pi&m_\nu+m_\pi-n+1
\end{array}\right)^2.
\label{e_sum2}
\end{equation}
\end{widetext}
The classical approximation consists of replacing
the last $3j$ symbol in Eq.~(\ref{e_sum2})
according to~\cite{Wigner59}
\begin{equation}
\left(\begin{array}{ccc}
J_\nu&J_\pi&J\\M_\nu&M_\pi&M
\end{array}\right)^2\mapsto
\left(4\pi A^{J_\nu J_\pi J}_{M_\nu M_\pi M}\right)^{-1},
\label{e_clas}
\end{equation}
where $A^{J_\nu J_\pi J}_{M_\nu M_\pi M}$
is related to the Caley determinant,
\begin{equation}
\left(A^{J_\nu J_\pi J}_{M_\nu M_\pi M}\right)^2=-\frac{1}{16}
\left|
\begin{array}{cccc}
0&a_{J_\nu M_\nu}&a_{J_\pi M_\pi}&1\\
a_{J_\nu M_\nu}&0&a_{JM}&1\\
a_{J_\pi M_\pi}&a_{JM}&0&1\\
1&1&1&0
\end{array}\right|,
\label{e_caley}
\end{equation}
with $a_{JM}\equiv(J+{\frac12})^2-M^2$.
While the $3j$ symbol in Eq.~(\ref{e_clas})
usually is a rapidly oscillating function of $M_\nu$ and $M_\pi$,
its classical approximation is smooth.
Many oscillations occur except when $J$ is close to its minimum or maximum value,
$J=|J_\nu-J_\pi|$ or $J=J_\nu+J_\pi$, respectively.
The approximation~(\ref{e_clas}) in the sum~(\ref{e_sum2}), therefore,
can be expected to be reasonable in most cases,
in particular in the region of physics interest where $J^2\approx J_\nu^2+J_\pi^2$.
Since $J_\rho\approx j_{1\rho}+j_{2\rho}$,
a classical approximation cannot be made
for the first two $3j$ symbols in Eq.~(\ref{e_sum2}).
From the explicit expressions for the $3j$ symbols
one can show that the sum over $M_\nu$ and $M_\pi$
can be restricted to the region around  $M_\nu=M_\pi=0$.
Since the classical approximation~(\ref{e_clas})
is nearly constant for small values of $M_\nu$ and $M_\pi$,
the classical limit for the $3j$ symbols
with zero projections on the $z$ axis can be taken,
to obtain
\begin{equation}
\sigma_0^{(0)}\approx \sigma_1^{(0)}\approx
\frac{(2J_\nu+1)(2J_\pi+1)}{4\pi(2j_{1\nu}+1)(2j_{1\pi}+1)A},
\label{e_sum3}
\end{equation}
where $A\equiv A^{J_\nu J_\pi J}_{000}$ is the area of a triangle with sides
of lengths $J_\nu+{\frac12}$, $J_\pi+{\frac12}$, and $J+{\frac12}$~\cite{note3}.
We may alternatively express this sum in terms of the shears angle,
\begin{equation}
\theta_{\nu\pi}=
\arccos\frac{J(J+1)-J_\nu(J_\nu+1)-J_\pi(J_\pi+1)}
{2\sqrt{J_\nu(J_\nu+1)J_\pi(J_\pi+1)}},
\end{equation}
leading to
\begin{equation}
\sigma_0^{(0)}\approx \sigma_1^{(0)}\approx
\frac{2}{\pi(2j_{1\nu}+1)(2j_{1\pi}+1)\sin\theta_{\nu\pi}}.
\label{e_sum4}
\end{equation}
A similar derivation yields the classical approximation
\begin{equation}
\sigma_0^{(1)}\approx
-(-)^{j_\nu+j_\pi}\frac{2}{\pi(2j_{1\nu}+1)(2j_{1\pi}+1)\tan\theta_{\nu\pi}}.
\label{e_sb0}
\end{equation}

Introducing the preceding approximations
into the expression for the shears matrix element~(\ref{e_result1}),
we obtain
\begin{equation}
\langle NP^{-1};J|V_{\nu\pi}^{\rm SDI}|NP^{-1};J\rangle\approx
\frac{s_2}{2\pi\sin\theta_{\nu\pi}}+
\frac{t_2}{2\pi\tan\theta_{\nu\pi}},
\label{e_sdi}
\end{equation}
where
\begin{eqnarray}
&&s_2=4(3a_0+a_1),
\qquad
t_2=\varphi(a_0-a_1),
\nonumber\\
&&\varphi={\cal P}(-)^{\ell_{1\nu}+j_{1\nu}+\ell_{1\pi}+j_{1\pi}}.
\end{eqnarray}
The classical limit of the shears matrix element
for the SDI does not depend
on the individual single-particle angular momenta
but only on the shears angle,
which is defined by the angular momenta $J_\nu$ and $J_\pi$ of the blades
and the total angular momentum $J$.

\begin{figure}
\includegraphics[width=7cm]{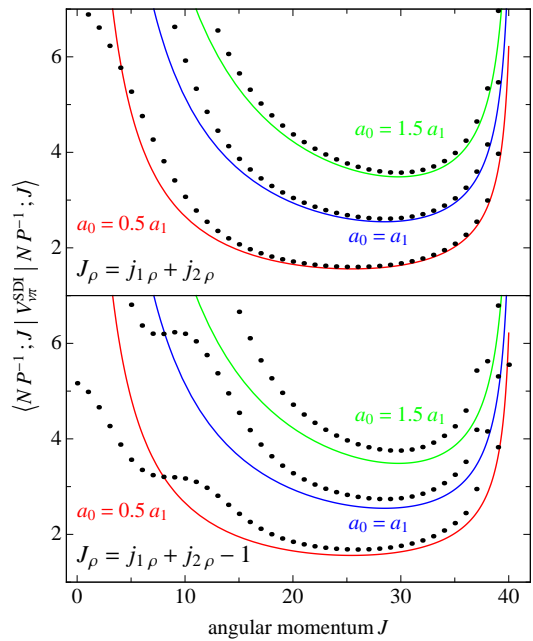}
\caption{
\label{f_2ph20}
The exact expression~(\ref{e_result1})
for the 2p-2h shears matrix element of the SDI (dots)
compared with its classical approximation~(\ref{e_sdi}) (lines)
for neutron and proton angular momenta $J_\nu=J_\pi=20$.
The single-particle angular momenta
are $j_{1\rho}=\frac{19}{2}$ and $j_{2\rho}=\frac{21}{2}$ (top),
and $j_{1\rho}=j_{2\rho}=\frac{21}{2}$ (bottom).
Results are shown for three choices of the ratio $a_0/a_1$.
The matrix element is in units $a_1$.}
\end{figure}

In Fig.~\ref{f_2ph20} the exact shears matrix elements of the SDI
are compared with their classical approximation
for the neutron and proton angular momenta of the blades $J_\nu=J_\pi=20$.
The approximation is good for the stretched case, $J_\rho=j_{1\rho}+j_{2\rho}$,
but deteriorates rapidly as $J_\rho$ diminishes.
However, in the region of interest,
where the two angular momenta $J_\nu$ and $J_\pi$ are close to orthogonal,
$J^2\approx J_\nu^2+J_\pi^2$,
the classical approximation is good,
also for $J_\rho=j_{1\rho}+j_{2\rho}-1$,
and the geometric picture underlying the shears mechanism remains valid.

\begin{figure}
\includegraphics[width=7cm]{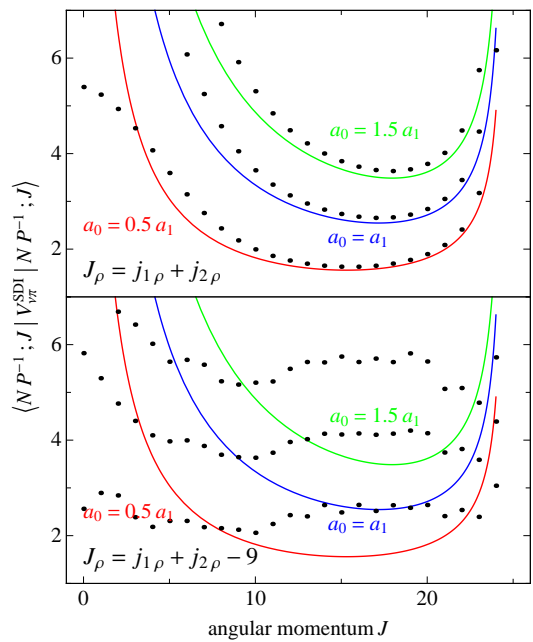}
\caption{
\label{f_2ph12}
Same as Fig.~\ref{f_2ph20} for neutron and proton angular momenta $J_\nu=J_\pi=12$
and single-particle angular momenta
$j_{1\rho}=\frac{11}{2}$ and $j_{2\rho}=\frac{13}{2}$ (top),
and $j_{1\rho}=j_{2\rho}=\frac{21}{2}$ (bottom).}
\end{figure}

This analysis also reveals the limits
of the validity of this geometric picture,
as is illustrated in Fig.~\ref{f_2ph12}
which shows the shears matrix element of the SDI
for the neutron and proton angular momenta $J_\nu=J_\pi=12$.
In one case these arise from aligned single-particle angular momenta
($j_{1\rho}=\frac{11}{2}$ and $j_{2\rho}=\frac{13}{2}$)
and the classical approximation is seen to be reasonable.
In the second case the single-particle angular momenta are larger ($j_{1\rho}=j_{2\rho}=\frac{21}{2}$)
and they are not aligned ($J_\rho=j_{1\rho}+j_{2\rho}-9$),
leading to a breakdown of the shears interpretation.
The alignment of the particles in high-$j$ orbits,
introduced here by hand,
is in a more realistic shell-model calculation
due to their interaction with particles in low-$j$ orbits~\cite{Frauendorf96}.

The classical expression~(\ref{e_sdi})
is reminiscent of the interpretation of nuclear matrix elements
in terms of the angle between the vectors of the single-particle angular momenta
(see, {\it e.g.}, Schiffer and True~\cite{Schiffer76}).
In fact, it can be shown that the classical limit of the 1p-1h matrix element
$\langle j_\nu j_\pi^{-1};J|V_{\nu\pi}^{\rm SDI}|j_\nu j_\pi^{-1};J\rangle$
leads to the same dependence
on the angle $\theta_{\nu\pi}$ as in Eq.~(\ref{e_sdi})
but with coefficients $s_1$ and $t_1$
that depend differently on the strengths $a_T$ of the interaction.

The present results make contact
with the semi-classical analysis~\cite{Clark00,Hubel05}
in terms of an interaction $V_0 + V_2P_2(\cos \theta_{\nu\pi})$,
assumed to exist between the blades of the shears bands.
In line with the intuitive picture one has of the shears mechanism,
this interaction necessarily implies a shears-band head
with $\cos\theta_{\nu\pi}^{\rm 0}\approx0$
or, alternatively, a minimum energy for $J^2\approx J_\nu^2+J_\pi^2$
when the angular momentum vectors $\bar J_\nu$ and $\bar J_\pi$
are orthogonal.
Our analysis shows that this is not generally valid
since a minimum energy is obtained from Eq.~(\ref{e_sdi}) for
$\cos\theta_{\nu\pi}^{\rm 0}\approx
-t_2/s_2=\varphi(a_1-a_0)/4(3a_0+a_1)$.
Therefore, if the isoscalar and isovector interaction strengths are different,
a minimum energy is found for a {\em non-orthogonal} configuration.

\begin{figure}
\includegraphics[width=6cm]{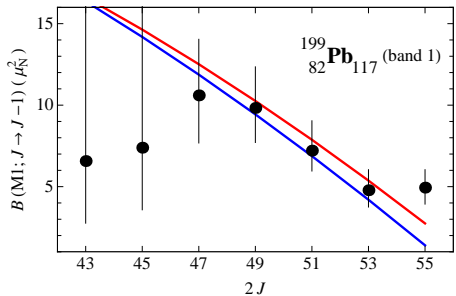}
\caption{
\label{f_pb199}
Experimental (points) and calculated (lines)
$B$(M1) values for band 1 in $^{199}$Pb
as a function of twice the angular momentum $J$.
The red line is the exact expression
and the blue line is its classical limit~(\ref{e_bm1}).}
\end{figure}

Shears bands are characterized by strong M1 transitions
that decrease with increasing angular momentum $J$.
For a general configuration $|NP^{-1};J\rangle$,
where $|N\rangle$ is an $m$-particle neutron
and $|P^{-1}\rangle$ an $m'$-hole proton configuration
with angular momenta $J_\nu$ and $J_\pi$, respectively,
standard recoupling techniques~\cite{Talmi93}
lead to the following classical expression for the $B$(M1) values:
\begin{eqnarray}
\lefteqn{B({\rm M1};J\rightarrow J-1)}
\nonumber\\&\approx&
\frac{3}{4\pi}(g_\nu-g_\pi)^2\frac{(2J_\nu+1)^2(2J_\pi+1)^2}{16J(2J+1)}\sin^2\theta_{\nu\pi},
\label{e_bm1}
\end{eqnarray}
where $g_\nu$ and $g_\pi$ are the $g$ factors
of the neutron and proton configurations
$|N\rangle$ and $|P^{-1}\rangle$, respectively.
This expression can be tested in $^{199}$Pb
where lifetimes of some shears-band levels have been measured~\cite{Neffgen95}.
In particular, the states in band~1 with $\frac{43}{2}<J<\frac{55}{2}$
have the suggested~\cite{Baldsiefen94} configuration
$\nu(1i_{13/2}^{-3})_{33/2}\times\pi(1h_{9/2}1i_{13/2})_{11}$,
from where the $g$ factors can be obtained,
$g_\nu=-0.29~\mu_{\rm N}$ and $g_\pi=1.03~\mu_{\rm N}$.
Application of Eq.~(\ref{e_bm1}) with $J_\nu=\frac{33}{2}$ and $J_\pi=11$
leads to the result shown in Fig.~\ref{f_pb199} (in blue),
which is seen to be close to the exact result (in red).
In the spin region $\frac{47}{2}\leq J\leq\frac{53}{2}$,
where the above configuration is thought to apply,
agreement is obtained.

The data on M1 transitions indicate that fixed neutron and proton configurations
do not apply to an entire shears band but only to part of it.
For this reason it will be difficult to use energy formulas like Eq.~(\ref{e_sdi})
with constant coefficients $s_2$ and $t_2$ for an entire band.
The present analysis suggests however
that, under certain conditions of angular-momentum alignment,
the generic form of the expression~(\ref{e_sdi}) is approximately valid
with coefficients $s_i$ and $t_i$ depending on the structure
of the neutron and proton configurations
that apply to certain ranges of the total angular momentum $J$.
This problem is currently under study.

This work was partially supported (AOM)
by the Director, Office of Science,
Office of Nuclear Physics,
of the U.S. Department of Energy
under contract No. DE-AC02-05CH11231
and by FUSTIPEN
(French-U.S. Theory Institute for Physics with Exotic Nuclei)
under DOE grant No. DE-FG02-10ER41700.


\begin{references}
\bibitem{note1}
The notation $j_\rho$ is used
as an abbreviation for the set of quantum numbers $n_\rho$, $\ell_\rho$, and $j_\rho$
of singe-particle levels in a central potential.
\end{references}
\end{document}